\documentclass[preprint,tightenlines,superscriptaddress,
prd,nofootinbib,eqsecnum,showpacs]{revtex4}

\usepackage{graphicx}
\def\beq{\begin{equation}} \def\eeq{\end{equation}}
\def\bea{\begin{eqnarray}} \def\eea{\end{eqnarray}}

\def\real{{\rm I\hskip-2pt R}}  \def\ts#1{\textstyle{#1}}
      
\def\ev#1{\langle #1 \rangle}   
\def\dd{{\rm d}}       \def\ee{{\rm e}}       \def\ii{{\rm i}}
      \def\eps{\epsilon}
\def\kin{_{{\rm kin}}} \def\half{{\textstyle{1\over2}}}
\def\phy{_{{\rm phy}}} \def\fourth{{\textstyle{1\over4}}}
\def\bb{\bar\beta} \def\pb{\bar p}
\def\tb{\tilde\beta} \def\tp{\tilde p}

\begin{document}
\preprint{\vbox{\baselineskip=12pt \rightline{ICN-UNAM-04/03}
\rightline{gr-qc/0404004} }}

\title{Semiclassical States in Quantum Cosmology:\\
Bianchi I Coherent States}

\author{Brett Bolen}\email{brett@phy.olemiss.edu}
\affiliation{Department of Physics and Astronomy,
University of Mississippi\\ University, MS 38677-1848, U.S.A.}
\affiliation{Department of Physics, Rhodes College\\
Memphis, TN 38112-1690, U.S.A.}
\author{Luca Bombelli}\email{luca@phy.olemiss.edu}
\affiliation{Department of Physics and Astronomy,
University of Mississippi\\ University, MS 38677-1848, U.S.A.}
\affiliation{Perimeter Institute for Theoretical Physics\\
35 King Street North, Waterloo, Ontario, Canada N2J 2W9}
\author{Alejandro Corichi}\email{corichi@nuclecu.unam.mx}
\affiliation{Department of Physics and Astronomy,
University of Mississippi\\ University, MS 38677-1848, U.S.A.}
\affiliation{Instituto de Ciencias Nucleares,
Universidad Nacional Aut\'onoma de M\'exico\\
Apartado Postal 70-543, M\'exico D.F. 04510, M\'exico\\ \ }

\date{9 June 2004}

\begin{abstract}
We study coherent states for Bianchi type I cosmological models,
as examples of semiclassical states for time-reparametrization
invariant systems. This simple model allows us to study explicitly
the relationship between exact semiclassical states in the
kinematical Hilbert space and corresponding ones in the physical
Hilbert space, which we construct here using the group averaging
technique. We find that it is possible to construct good
semiclassical physical states by such a procedure in this model;
we also discuss the sense in which the original kinematical states
may be a good approximation to the physical ones, and the
situations in which this is the case. In addition, these models
can be deparametrized in a natural way, and we study the effect of
time evolution on an ``intrinsic" coherent state in the reduced
phase space, in order to estimate the time for this state to
spread significantly.
\end{abstract}
 \maketitle

\section{Introduction}

\noindent Bianchi cosmological models have long been used as toy
models in both classical general relativity and quantum gravity. The
small but non-trivial number of degrees of freedom they possess makes
their dynamics interesting from the point of view of shedding light
on features of the full theory, even though it reduces to that of a
relativistic particle in a low-dimensional space. These models are
homogeneous cosmologies, in which the symmetries are imposed on the
canonical action or Hamiltonian before the dynamics is obtained. This
procedure is, in general, not equivalent to imposing the symmetries
on the full dynamical equations, but in a classical theory this can
only potentially give some extra solutions with respect to the
symmetric ones for the full theory, and is not a problem
\cite{torre1}. Thus, the classical Bianchi models can be very
fruitfully used in exploring issues such as the evolution of
anisotropies in simplified versions of the early universe, or the
chaotic nature of the evolution of the metric close to the initial
singularity.

In the quantum theory, the difference between any physical
prediction of a minisuperpace model and the ``full theory" may be
more substantial \cite{mike}, but it is generally believed that some
qualitative features of the models are still useful indications of
what to expect from the full theory when different approaches to its
quantization are used. In particular, minisuperspace models are
time-reparametrization invariant, thus exhibiting the much
celebrated ``problem of time" \cite{time}; they represent then a
fruitful playground in which to explore quantization methods, in the
hope of getting some insight on how to approach this problem in full
quantum
gravity. It is not surprising then that minisuperpaces have been
analyzed from the Euclidean path integral approach to quantum
gravity \cite{EQG}, consistent histories \cite{CH} and refined
algebraic quantization \cite{RAQ} perspectives.

It is in this spirit that we study here coherent states of
minisuperspace models. The aim is to understand the role of coherent
states in defining semiclassical states for time-reparametrization
invariant systems.  In recent years, an increasing amount of work has
been devoted to understanding the phenomenology of semiclassical
states in quantum gravity, by which here we mean that sector of the
physical states in which the gravitational degrees of freedom are
sharply peaked around values satisfying the classical constraint
equations, because it is expected that these states will possess
quasi-classical properties in certain regimes \cite{abc}. The study
of the semiclassical limit of quantum cosmological models is of
course not new; however, most of the treatments that we are aware of
deal with the WKB method, where the wave function is assumed to be of
a special form and some component of it is shown to satisfy the
Hamilton-Jacobi equation (for a review, see Ref \cite{kiefer1}).

In this paper, for simplicity, we consider vacuum Bianchi type I models,
whose classical description can be reduced to that of a relativistic
particle moving in a 3-dimensional, linear configuration space,
subject to the Hamiltonian constraint of general relativity, which in
this case reduces to the free particle Hamiltonian. These models have
been extensively studied, but the type of questions we are concerned
with here, however, have been only recently considered within the
context of constrained systems in Ref \cite{abc}, where a general
framework for the construction of semiclassical states for a class of
constrained theories is discussed. Here, we follow much of that
approach in spirit, but the constraint and physical observables that
arise in these Bianchi models are not of the same type and we have to
adapt the general constructions to the present situation, and the
fact that we are dealing with cosmological models allows us to ask
additional physically interesting questions.

Despite the simplicity of the classical model, the quantum counterpart
is non-trivial, because of the presence of the constraint. First,
we consider coherent states in the kinematical Hilbert space, peaked
around a point on the constraint surface in phase space but without
taking the quantum constraint into account, and characterized by a
set of ``squeezing" parameters. These states are only meant to be
approximations to physical coherent states, i.e., solutions to
the quantum constraint, which we find by applying the group averaging
procedure of the Refined Algebraic Quantization program (RAQ)
\cite{RAQ} to the kinematical coherent states; one expects these
to be the ones that capture the ``essential" features of semiclassical
Bianchi I models. Finally, we consider the reduced phase space,
described by true Dirac observables, and define coherent states peaked
around reduced phase space points; one might call these ``intrinsic"
coherent states. The main purpose of this article is to compare these
three classes of states, in terms of describing semiclassical
situations, assess how good the kinematical coherent states are as
approximations to the physical ones, and use the ``intrinsic" ones to
get an insight into evolution issues.

The structure of the paper is the following. In Sec.~\ref{sec:2}
we recall some basic notions about Bianchi models. In
Sec.~\ref{sec:3} we recall the notion of  coherent states, and
discuss what we mean by semiclassical states for a system.
Sec.~\ref{sec:4} considers the kinematical coherent states for
Bianchi I models, and in Sec.~\ref{sec:5} the Physical coherent
states. In Sec.~\ref{sec:6} we study the time evolution of the
``intrinsic coherent states," and in particular some estimates of
the time it would take for a coherent state to lose its peakedness
properties. We end with a discussion in Sec.~\ref{sec:7}.
Throughout the paper we set $c = G = 1$,
but keep $\hbar$ dimensionful.

\section{Preliminaries on Bianchi Models and Their Observables}
\label{sec:2}

\noindent In this section, we will provide some background
information on Bianchi models, the variables used to describe them,
and their physical observables; readers who are familiar with these
models should be able to skim through it quickly.

Bianchi models are spatially homogeneous cosmologies: they describe
spacetimes which can be foliated by spacelike hypersurfaces
$\Sigma_t$ (all diffeomorphic to the ``space manifold" $\Sigma$), on
each of which a group $G$ of isometries acts transitively, and they
are classified by the Lie algebra of the group $G$.

If one uses three linearly independent left-invariant forms
$\sigma^i$ on $\Sigma$ as a basis for its cotangent space, the
spacetime metric can be written in the form $g = -N^2(t)\,\dd t
\otimes\dd t + q_{ij}(t)\, \sigma^i \otimes \sigma^j$, where in some
Bianchi models the matrix $g_{ij}$ can be diagonalized and written,
in the Misner parametrization \cite{misner}, as
\beq
    q_{ij} = \ee^{2\beta^0}\,{\rm diag}
    (\ee^{2(\beta^++\sqrt3\beta^-)},
    \ee^{2(\beta^+-\sqrt3\beta^-)},\ee^{-4\beta^+})\;;
    \label{metric}
\eeq
we thus see that $\sqrt{\det q} = \ee^{3\beta^0}$, and
$\beta^0(t)$ characterizes the volume of $\Sigma_t$, while the
$\beta^\pm(t)$ characterize its anisotropy; all three functions
take values in $(-\infty,+\infty)$. We will take as configuration
variables the $\beta^i$, with $i = 0,+,-$; the configuration space
is thus ${\cal C} = \real^3$. The dynamics of these models then
reduces (before imposing the constraints) to that of a
relativistic particle in 3 dimensions, and is governed by a
Hamiltonian constraint of the form
\beq
    H = \half\,\eta^{ij}p_ip_j + U(\beta^i)\;, \label{H}
\eeq
in the vacuum case,
where $p_i$ is the canonical momentum conjugate to $\beta^i$,
and $\eta_{ij}$ a flat metric on $\cal C$, with line element
$\eta_{ij}\, \dd\beta^i \dd\beta^j = -(\dd\beta^0)^2 +
(\dd\beta^+)^2 + (\dd\beta^-)^2$, and the form of the potential
$U$ depends on the model under consideration. Here we should point
out that even when the flat metric resembles that of 3D Minkowski
spacetime, it is actually conformally related to the De~Witt
supermetric on ${\cal C}$. As with all general relativistic
models, the Hamiltonian is also a constraint, in that allowed
classical phase space points have to satisfy the scalar constraint
$H(p_i,\beta^i) = 0$; this is the only constraint for these
models, since the metric (\ref{metric}) satisfies the vector
constraint identically.

In this paper, we restrict ourselves to begin with to the simplest
models, the vacuum ones of type I (with homogeneity group U(1)$^3$,
or the additive $\real^3$). In this case, $U(\beta^i)$ vanishes
identically; this allows us to not only solve the quantum constraint
easily, but also to conceptually isolate the issue of comparing the
different coherent states on a linear phase space with a relatively
simple constraint, from those related to the structure of phase
space or the form of the potential. In the canonical quantum theory,
following the Dirac prescription, states have to be annihilated by
the operator version $\hat H$ of the constraint, i.e., be invariant
under the group $\ee^{-\ii\lambda\hat H}$ it generates, and it is
therefore very advantageous, to use the $p$-representation, because
the Hamiltonian is then just a multiplication operator. We will
also rescale the variables by $p_i \mapsto p_i/\hbar$, which makes
the new $p_i$ dimensionless (like the $\beta^i$); we can think of this
alternatively as a rescaling of the action or Hamiltonian, or as a
way to give the value of $p_i$ in terms of the quantum unit $\hbar$,
which is useful when characterizing the extent to which a fluctuation
is small or large in the classical sense. For simplicity of notation,
in the rest of the paper we will use only subscripts, and denote
$\beta_i \equiv \beta^i$ (that is, the index is not lowered with the
metric $\eta_{ij}$).

In order to evaluate candidate semiclassical states by comparing
fluctuations of observables, we need to identify the proper set of
physical observables to consider. It is natural to look for a
sufficient number of Dirac observables, in the sense that their
values constitute a full characterization of a (classical,
gauge-invariant) state of the system. Since we will be using the
$p$-representation, to simplify the terminology it is convenient
to call the space $\tilde{\cal C} = \real^3$ of $p_i$'s our new
configuration space. Classically, configuration observables are
then functions $f(p_i)$ on $\tilde{\cal C}$, and ``momentum"
observables, specifically linear functions of the canonically
conjugate $\beta_i$, can be thought of as vector fields $u$ on
$\tilde{\cal C}$, since there is a canonical assignment of a
linear momentum observable $P_u$ everywhere in the phase space
$\Gamma$ to each such vector field; we can then associate to
these observables quantum operators $\hat f$ and $\hat u$ acting
on $\psi(p)$, defined respectively by
\beq
    (\hat f\,\psi)(p):= f(p)\,\psi(p)\;,\qquad
    (\hat u\,\psi)(p):= \ii\,({\cal L}_u
    + \ts{\frac12}\, {\rm div}_\mu u)\,\psi(p)\;, \label{operators}
\eeq
where div$_\mu u$ is the divergence of $u$ with respect to some volume
element $\dd\mu(p)$ on $\tilde{\cal C}$.

We know that the reduced phase space has four dimensions, so we
need four independent observables $\{O_\alpha\}_{\alpha=1,...,4}$,
in addition to the constraint itself. The first observation is that
the variables $p_\pm$ are Dirac observables, since the Hamiltonian
constraint in the new variables,
\beq
    H(\beta_i,p_i) = \half\,(-p_0^2 + p_+^2 + p_-^2)\;,
    \label{Hnew}
\eeq
depends on the $p_i$ only. The not so obvious task is that of finding
a suitable pair of linear momentum observables, or vector fields on
$\tilde{\cal C}$, to represent the conjugate variables, since the
$\beta_i$, for which the CCR with the $p_i$ are of the form
$[\hat\beta_i,\hat p_j] = \ii\, \delta^i{}_j$, are not physical
observables; in fact, one can easily see that no function of the
$\beta_i$ alone is a constant of the motion. As Dirac ``momentum"
observables for this model we will use instead the combinations
$v_\pm:= p_0\beta_\pm + p_\pm\beta_0$ which, together with
$v_0:= p_+\beta_- - p_-\beta_+$ are associated with the vector
fields (denoted here by the same symbols)
\beq
    v_{+} = p_0\,\frac{\partial}{\partial p_+}
    + p_+\,\frac{\partial}{\partial p_0}\;, \qquad
    v_{-} = p_0\,\frac{\partial}{\partial p_-}
    + p_-\,\frac{\partial}{\partial p_0}\;, \qquad
    v_0 = p_+\,\frac{\partial}{\partial p_-}
    - p_-\,\frac{\partial}{\partial p_+}\;,
\eeq
i.e., the two boosts and the spatial rotation that generate the (2+1)
Lorentz group \cite{ATU} on the `minisuperspace' $\tilde{\cal C}$.

We may note that in general there is a possible ambiguity in the
class of observables, consisting in adding to each one a term
proportional to the constraint; such an addition does not change the
value of the observable on the constraint surface. One can implement
a general strategy to remove the ambiguity \cite{abc}, but in the
present case no such change would preserve the property that all
observables are linear in the configuration
variables.

\section{Semiclassical and Coherent States}
\label{sec:3}

\noindent In this section, we will specify our criteria for a
quantum state to be semiclassical, following the general
discussion of Ref \cite{abc}, recall the basic definition of the
most common example of such a state, a coherent state, and specify
the setup and goals for the rest of the paper, specializing the
discussion to Bianchi I cosmology. (For other references on
semiclassical states for systems with constraints, see also Refs
\cite{Ashworth,DateSingh}.)

Given a point $(\bar q_i, \pb_i)$ in a classical phase space, a
complete set of physical observables
$\{O_\alpha\}_{\alpha=1,...,4}$ for the theory (completeness will
be the only requirement, the observables do not need to generate
an algebra; they may be overcomplete), and a pair of parameter
values $(\eps_\alpha, \delta_\alpha)$ for each $O_\alpha$, we will
consider a state $\psi$ to be semiclassical if for each $\alpha$
\beq
    |\ev{\hat O_\alpha}_\psi-O_\alpha(\bar q_i, \pb_i)|
    < \eps_\alpha\;,\qquad
    (\Delta\hat O_\alpha)_\psi < \delta_\alpha\;. \label{semi}
\eeq
In other words, we are assigning tolerances; the
$(\eps_\alpha, \delta_\alpha)$ give a measure of how close the
expectation values must be to the classical values, and how
narrowly peaked a state must be, for the latter to be considered
semiclassical. One often takes the $O_\alpha$ to be the canonical
variables $(\bar q_i, \pb_i)$ themselves, but this is not
necessary; in some cases, as we will see, one is forced to make
other choices, and in general the choice can affect in a crucial
way which states are considered to be semiclassical.

If the system has a set of constraints $H_I(q_i,p_i) = 0$ (assumed
here to be first class), a couple of additional points need to be
addressed. First of all, the observables need to be physical in
the Dirac sense of commuting (weakly) with the constraints,
$\{O_\alpha,H_I\} \approx 0$, but they need only be complete in
the reduced phase space---if one knows how to characterize it.
Secondly, if the state does not belong to the physical Hilbert
space, we specify the value of an additional pair of tolerance
parameters $(\eps_I,\delta_I)$ for each constraint, and require
that $\forall I$
\beq
    |\ev{\hat H_I}_\psi| < \eps_\alpha\;,\qquad
    (\Delta\hat H_I)_\psi < \delta_I\;.
\eeq

An important issue, which in fact underlies the motivation for the
current work, arises when one has a constrained system for which
one prefers, or is forced, to work with kinematical states. One
may want to know in that case whether the kinematical
semiclassical states are, in an appropriate sense, good
approximations to the physical ones, so that one may extract from
the former useful information on the latter, for example regarding
quantum fluctuations. The general formulation of this issue
deserves further study; here we will just mention that two
possible definitions of when $\psi\kin$ is a good approximation to
$\psi\phy$ are: (i) The two states are semiclassical with respect
to the same $(\bar q_i, \pb_i)$, with the same tolerances; and the
more restrictive definition (ii) For all $O_\alpha$
\beq
    {|\ev{\hat O_\alpha}\phy-\ev{\hat O_\alpha}\kin|
    \over|\ev{\hat O_\alpha}\phy|}\ ,
    \ {|(\Delta\hat O_\alpha)^2\phy-(\Delta\hat O_\alpha)^2\kin|
    \over(\Delta\hat O_\alpha)^2\phy}\ll1\;.
\eeq
Qualitatively, the first definition simply aims at
identifying situations in which the spreads of measured values for
the $O_\alpha$ in $\psi\kin$ and $\psi\phy$ are both within the
same tolerances, while the second one may be suited for situations
in which one would like all quantities which can be calculated
from the spread of values of $O_\alpha$ with the two states to
nearly coincide.

It would be natural to ask at this point whether the values of the tolerances $(\eps_\alpha,\delta_\alpha)$ can be arbitrarily chosen. Although the definition above is consistent for arbitrary (positive) values, if the $\delta_\alpha$ are too small no state will satisfy the conditions. For example, in the case of a pair of conjugate variables $(p,q)$, choosing $\delta_p$ and $\delta_q$ such that $\delta_p\,\delta_q < \hbar/2$ would be too restrictive; in general, quantum uncertainty relations will translate into some (macroscopically) mild conditions that set lower bounds for combinations of $\delta_\alpha$'s. The $\eps_\alpha$ in principle can be arbitrary, but in practice they may be restricted by the way the states are constructed, as mentioned below, and by their interpretation. A physically motivated choice would be to set the value of each $\eps_\alpha$ to be close to, or smaller than, that of the corresponding $\delta_\alpha$, with the latter being smaller than the experimental resolution for $O_\alpha$.

We should also comment here that it may seem like the introduction of the
parameters $\eps_\alpha$ is unnecessary, since one can usually choose the
states in such a way that the expectation values of the $\hat O_\alpha$
agree with their classical values. This is correct, in principle; however,
some of the physical states we will use are constructed (using group averaging) out of states with the right expectation values for our choice of physical observables, but do not have this property themselves. Besides, it may sometimes simplify calculations to choose states which don't have exactly the classical expectation values for the chosen $\hat O_\alpha$.

Coherent states are among the most commonly used states with a
semiclassical interpretation; they are relatively easy to work with, are often associated with realizable physical situations, and satisfy minimum uncertainty relations. However, those relations only refer to the product of uncertainties for pairs of conjugate variables, and not individual
uncertainties, so given some set of $\{(O_\alpha, \eps_\alpha,
\delta_\alpha)\}_{\alpha=1,...,4}$, not all coherent states may satisfy
our definition of semiclassical states. To look at this question, let us
summarize the freedom available in defining coherent states for a theory.

Consider a theory with a linear phase space $\Gamma = {\bf R}^{2N} =
\{(q_i,p_i)\}_{i=1,...N}$, where the phase space coordinates are
dimensionless. Then a coherent state centered at $(\bar q_i, \bar
p_i) \in \Gamma$ in the $p$-representation is (the square root of) a
Gaussian wave function \cite{Schiff},
\beq
    \psi_{\bar p,\bar q,\sigma}(p) = \prod\nolimits_{i=1}^N
    {\cal N}_i\,\exp\left\{ -(p_i-\bar p_i)^2/4\sigma_i^2
    - \ii\,\bar q_i\,(p_i-\bar p_i) \right\}, \label{coh}
\eeq
which depends on the $N$ parameters $\sigma_i$, the widths of the
Gaussian in the various $p_i$ directions; ${\cal N}_i = (\sqrt{2\pi}
\sigma_i)^{-1/2}$ is a normalization factor. The values of the $\sigma_i$
are the freedom we have (once we fix the classical phase space point
$(\bar q_i,\bar p_i)$) to ensure that the state is semiclassical.
Since the Fourier transform of a Gaussian of width $\sigma_i$ is
another Gaussian, of width $1/2\sigma_i$, we expect that for
each choice of observables and tolerances $\{(O_\alpha, \eps_\alpha,
\delta_\alpha)\}$ there will be a finite range of values of the
$\sigma_i$ for which the state $\psi_{\bar p,\bar q,\sigma}$ is
semiclassical, provided the tolerances are large enough to be consistent
with the uncertainty relations.

In the familiar case of a simple harmonic oscillator, where
the Hamiltonian contains a pair of parameters $(m_i,\omega_i)$ for each
degree of freedom, one normally takes $\sigma_i = \sqrt{m_i \omega_i
\hbar/2}$, and ends up with the coherent states commonly used in that
case, while other choices for the $\sigma_i$ lead to squeezed states.
Here, we don't have such dimensionful parameters available, and it
might seem that this fact will make the process of defining coherent
states somewhat difficult. In reality, this is not the case; we don't
have the same natural choice for the $\sigma_i$, but we have instead
the freedom of choosing values that will make our states satisfy our
semiclassicality conditions.

The situation for Bianchi I models is then as follows. Our
observables $O_\alpha$ are the $p_\pm$ and $v_\pm$, and we have a
single constraint $H$; since the two quantities in each of the two
observable pairs have a similar physical interpretation, we will
assign the same tolerances to them. Thus, given a classical state
$(\pb_i, \bb_i)\in \bar \Gamma$ and a set of tolerance parameters
$(\eps_p, \eps_v, \delta_p, \delta_v, \eps_H, \delta_H)$, a
quantum state $\psi$ will be called semiclassical if it satisfies
the conditions
\beq
    |\ev{\hat p_\pm}_\psi-\pb_\pm| < \eps_p\;, \qquad
    |\ev{\hat v_\pm}_\psi-\bar v_\pm| < \eps_v\;, \qquad
    (\Delta\hat p_\pm)_\psi < \delta_p\;, \qquad
    (\Delta\hat v_\pm)_\psi < \delta_v\;, \label{semi1}
\eeq
and, if $\psi$ is not a solution of the quantum constraint,
\beq
    |\ev{\hat H}_\psi| < \eps_H\;, \qquad
    (\Delta\hat H)_\psi < \delta_H\;. \label{semi2}
\eeq
Notice that we could have chosen the observables $p_0$ and
$v_0$ as part of our set, and assigned tolerances to them as well.
In this model, our results would not have been very different,
although in general one cannot even assume that observables which
are functions of other operators that have small quantum
fluctuations have small fluctuations themselves.

In the kinematical Hilbert space, based on a linear configuration
space, we can use coherent states of the form (\ref{coh}). Then,
given a set of tolerances we will have more inequalities than free
parameters (the $\sigma_i$ or a set of equivalent ones); although
it is reasonable to expect that solutions do exist for
sufficiently large $(\eps_\alpha, \delta_\alpha)$, one of our
goals is to check that this is true for ``small" tolerances as
well. In the physical Hilbert space these states cannot be used
directly, but we will follow the group averaging procedure to set
up appropriately modified coherent states peaked around points of
the reduced phase space, check if they are semiclassical as well,
and whether the original kinematical states can be considered good
approximations. In the concluding section, we will mention models
for which the configuration space is not linear, in which case one
needs to replace the coherent states defined above by a different
choice. The specific form of the states used will depend on the
characteristics of the model, and we will discuss some options
later.

\section{Kinematical Coherent States}
\label{sec:4}

\noindent The first step in the quantization procedure is to find  a
representation of the CCR on the relevant Hilbert space. In our case,
given the form of the Hamiltonian, the most convenient choice for the
kinematical Hilbert space is to set ${\cal H}_{\rm kin} = {\rm L}^2
(\tilde{\cal C}, \dd p_j)$. In this representation, the $\hat p_j$
operators act by multiplication and the $\hat\beta_i$ act as
$\hat\beta_i = \ii\, \partial/\partial p_i$; this is the standard
Schr\"odinger representation of the CCR. As for the additional
observables $v_i$, since the measure $\mu$ on ${\cal H}\kin$ is
trivial, we simply get div$_\mu\,v_\pm$ = div$_\mu\,v_0 = 0$, and
\beq
    \hat v_\pm\psi = \ii\,(p_0\,\partial_\pm+p_\pm\,\partial_0)\,\psi\;,
    \qquad\hat v_0\psi = \ii\,(p_+\,\partial_- - p_-\,\partial_+)\,\psi
    \;,
\eeq
where $\partial_i:= \partial/\partial p_i$.

Given a point $\bar P = (\bb_i,\pb_i)$ on the constraint surface
$\bar\Gamma$ in the classical phase space $\Gamma$, we construct a
coherent state centered at $\bar P$ in the above representation with the
usual prescription (\ref{coh}), as
\beq
    \psi\kin(p_i) = \prod\nolimits_{i\,=\,0,\pm}
    {\cal N}_i\exp\left\{- (p_i-\pb_i)^2/2\tau_i
    - \ii\,(p_i-\pb_i)\bb_i \right\}, \label{psikin}
\eeq
where the free parameters $\tau_i$ measure the width of the coherent
state in momentum space (in the natural Planck units), analogously to the
$\sigma_i$ in (\ref{coh}), and the normalization factors are ${\cal N}_i =
(\pi\tau_i)^{-1/4}$. In this state, we recover the usual properties of
coherent states, namely that the expectation values of the canonical
variables are
\beq
    \ev{\hat\beta_i}\kin = \bb_i\;, \qquad
    \ev{\hat{p}_i}\kin = \pb_i\;,
\eeq
and their fluctuations have the following expressions
\beq
    (\Delta \hat\beta_i)^2\kin = 1/2\tau_i\;,\qquad
    (\Delta \hat{p}_i)^2\kin = \tau_i/2\;,
\eeq
so that $(\Delta \hat\beta_i)(\Delta \hat{p}_i) = \frac12$ for each $i$
(no summation over $i$). In practice, given the similar interpretation
of $\beta_+$ and $\beta_-$, we will often set $\tau_+ = \tau_-$, for
simplicity. For the moment, we shall leave the three $\tau_i$ unspecified,
but we will assume that the state is narrowly peaked in the sense that
the $\tau_i$ are small, $\tau_i/\pb_0^2 \ll 1$ for all $i$.

Although we are not imposing the quantum constraint on this state,
since we will compare it with a corresponding physical coherent
state, the quantities we need to calculate are the expectation
values and fluctuations of the physical operators we have identified,
in the state $\psi\kin$. We already have those of $\hat p_\pm$; for
the remaining ones (among which we include $v_0$, although strictly
speaking it is not necessary), the expectation values are
\bea
    & &\ev{\hat v_\pm}\kin = \int\dd^3p\;\psi^*\kin(p_i)
    \,\ii \left(p_0\,\partial_\pm + p_\pm\,\partial_0 \right)\psi\kin(p_i)
    = \pb_0 \bb_\pm + \pb_\pm \bb_0 \;, \\
    & &\ev{\hat v_0}\kin = \int\dd^3p\;\psi^*\kin(p_i)
    \,\ii \left(p_+\,\partial_- - p_-\,\partial_+ \right)\psi\kin(p_i)
    = \pb_+ \bb_- - \pb_- \bb_+ \;,
\eea
and for the fluctuations $(\Delta \hat O_\alpha)\kin^2 = \ev{\hat
O_\alpha^2}\kin - \ev{\hat O_\alpha}\kin^2$ we obtain
\bea
    & &(\Delta \hat v_\pm)\kin^2
    = \half + \fourth \left({\tau_0\over\tau_\pm}
    + {\tau_\pm\over\tau_0} \right)
    + \half \left( {\pb_\pm^2\over\tau_0}
    + {\pb_0^2\over\tau_\pm} \right)
    + \half\,(\bb_0^2\tau_\pm + \bb_\pm^2\tau_0)
    \;,\\ & &(\Delta \hat v_0)\kin^2
    = -\half + \fourth \left({\tau_+\over\tau_-}
    + {\tau_-\over\tau_+} \right)
    + \half \left( {\pb_+^2\over\tau_-}
    + {\pb_-^2\over\tau_+} \right)
    + \half\,(\bb_-^2\tau_+ + \bb_+^2\tau_-) \;.
\eea
In addition, for the Hamiltonian constraint operator we get
\bea
    & &\ev{\hat H}\kin = \fourth\,(-\tau_0 + \tau_+ + \tau_-)
    \label{Hkin} \\
    & &(\Delta\hat H)\kin^2
    = \half\,(\pb_0^2\tau_0 + \pb_+^2\tau_+ + \pb_-^2\tau_-)
    + {\ts{1\over8}}\,(\tau_0^2 + \tau_+^2 + \tau_-^2)\;,
\eea
where we have used the classical constraint $-\pb_0^2 + \pb_+^2 + \pb_-^2
= 0$. Thus, as expected on general grounds \cite{abc}, the
expectation values of all relevant operators agree with the classical
values, except for the (quadratic) Hamiltonian. In addition, for the
kinematical coherent state to be a good semiclassical state according
to our choice of tolerances, the above fluctuations need to satisfy
the conditions (\ref{semi1}) and (\ref{semi2}).

Specifically, the only nontrivial condition arising from the expectation
values of observables is the one from the constraint in (\ref{semi2}),
\beq
    |\tau_++\tau_--\tau_0| < 4\,\eps_H\;, \label{epsH}
\eeq
and the conditions on the uncertainties in the $\hat p_\pm$, the
$\hat v_\pm$ and $\hat H$ give, respectively
\bea
    & &\tau_\pm < 2\,\delta_p^2 \label{deltap} \\
    & &\half + \fourth \left({\tau_0\over\tau_\pm}
    + {\tau_\pm\over\tau_0} \right)
    + \half \left( {\pb_\pm^2\over\tau_0}
    + {\pb_0^2\over\tau_\pm} \right)
    + \half\,(\bb_0^2\tau_\pm + \bb_\pm^2\tau_0) < \delta_v^2
    \label{ineq2} \\
    & &\half\,(\pb_0^2\tau_0 + \pb_+^2\tau_+ + \pb_-^2\tau_-)
    + {\ts{1\over8}}\,(\tau_0^2 + \tau_+^2 + \tau_-^2) < \delta_H^2
    \label{ineq1}
\eea

These inequalities could certainly always be satisfied if all
tolerances were considered as free parameters, for example by
choosing large enough values of the $\delta_\alpha$, but what we
are interested in is finding, for fixed
$\{(\eps_\alpha,\delta_\alpha)\}$, a range of values for the
$\tau_i$ which satisfy the inequalities. We will show explicitly
how to do this in the case $\tau_+ = \tau_- =: \tau$, which
simplifies both the calculations and the visualization of the
results.

\begin{figure}
  \includegraphics[angle=0,scale=.70]{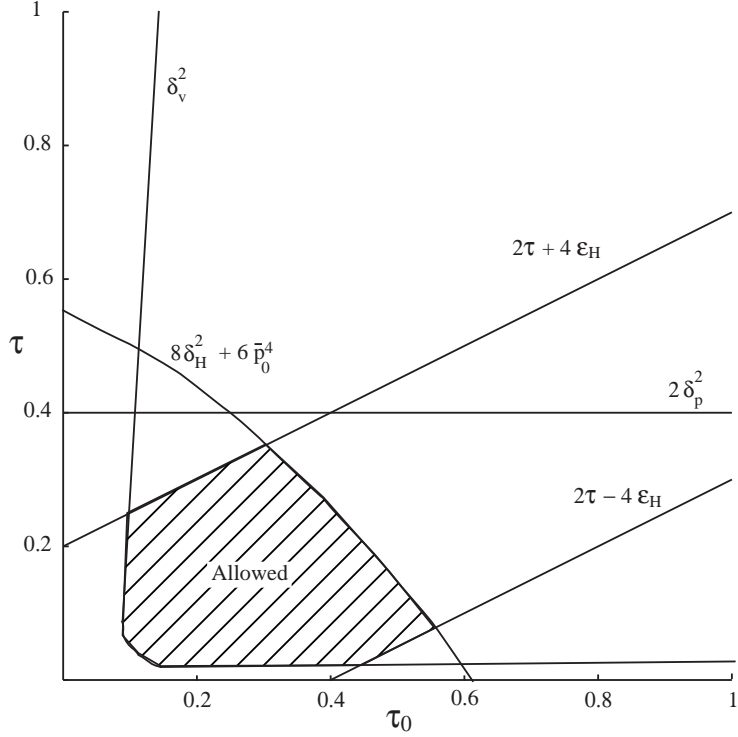}
  \caption{\label{f1}
The region in the parameter space of $(\tau_0,\tau)$ where the
inequalities for a semiclassical state are satisfied is shown
together with several of the curves that bound the region.}
\end{figure}

For example, Eq (\ref{ineq1}) becomes now a simple quadratic relation
between $\tau$ and $\tau_0$,
\beq
    \tau_0^2 + 4\,\pb_0^2\tau_0 +
    (4\,\pb_0^2\tau + 2\,\tau^2 - 8\,\delta_H^2) < 0\;, \label{28}
\eeq
which can also be written in a form showing explicitly that
$(\tau,\tau_0)$ must be inside an ellipse centered at $(-\pb_0^2,
-2\pb_0^2)$. The full set of inequalities is
\bea
    & &0 < \tau < 2\,\delta_p^2 \label{one} \\
    & &2\tau-4\eps_H < \tau_0 < 2\tau + 4\eps_H \vphantom{\sqrt f}
    \label{three} \\
    & &(\tau_0 + 2\,\pb_0^2)^2 + 2\,(\tau+\pb_0^2)^2
    < 8\,\delta_H^2 + 6\,\pb_0^4 \vphantom{\sqrt f} \label{deltaH} \\
    & &(\bb_0^2\tau + \bb_\pm^2\tau_0)\,\tau_0\tau
    + (1 - 2\,\delta_v^2)\,\tau_0\tau
    + \half\, (\tau_0^2 + \tau^2) + (\pb_\pm^2\tau + \pb_0^2\tau_0)
    < 0\;, \vphantom{\sqrt f}\qquad \label{deltav}
\eea
where the last inequality is equivalent to (\ref{ineq2}) for
non-vanishing $\tau$ and $\tau_0$.

Fig.~\ref{f1} illustrates roughly the range of allowed
$(\tau,\tau_0)$ values, for the choice of parameter values $\pb_0
= 1$, $\pb_\pm = 1/ \sqrt2$, $\bb_0 = 1$, $\bb_\pm = 1/ \sqrt2$,
$\delta_p = 0.2$, $\delta_v = 10$, $\epsilon_H = 0.1$, $\delta_H =
0.6$. Strictly speaking, many values in the ``allowed region"
shown do not satisfy the conditions under which our
small-$\tau/\pb_0^2$ approximation is valid, but the plot gives a
useful qualitative description of the effect of each inequality in
the $\tau_0$-$\tau$ plane.

The following argument tells us that the first three inequalities
can in practice be replaced by one of them. Notice that, given
that we are assuming $\pb_0$ to be large, and the relationship
between $H$ and the $p_i$'s, it is reasonable to set for
simplicity
\beq
    \eps_H \approx \delta_H \approx \pb_0\,\delta_p\;,\qquad
    {\rm so}\qquad \delta_p^2 \ll \eps_H,\quad\,\quad\delta_H
    \ll \pb_0^2\;.
\eeq
Consider first the inequality (\ref{three}). It defines a
strip of allowed $(\tau,\tau_0)$ values around the $\tau_0 =
2\,\tau$ line, and this strip intersects the $\tau$ and $\tau_0$
axes, respectively, at
\beq
    \tau = 2\,\eps_H\;,\qquad \tau_0 = 4\,\eps_H\;.
\eeq
With the choice of parameters we just made, $2\,\eps_H >
2\,\delta_p^2$, so the first of the two inequalities in
(\ref{three}) becomes redundant, provided that (\ref{one}) is
satisfied. However, if we now look at (\ref{deltaH}), it is easy
to see that the ellipse intersects the axes at, respectively,
\beq
    \tau = \sqrt{\pb_0^4 + 4\,\delta_H^2} - \pb_0^2 \approx
    2\,\delta_p^2\;, \qquad
    \tau_0 = \sqrt{4\,\pb_0^4 + 8\,\delta_H^2} - 2\,\pb_0^2
    < 4\,\eps_H\;,
\eeq
where we have used the fact that $\pb_0$ is large in the inequalities.
Thus, kinematical coherent states are semiclassical if $(\tau,\tau_0)$
lie inside the portion of the ellipse (\ref{deltaH}) that defines a
neighborhood of the origin in the first quadrant,
\beq
    0 < \tau_0 < 2\,\pb_0^2
    \left(\sqrt{1 + {2\,\delta_H^2\over\pb_0^4}} - 1\right), \qquad
    0 < \tau < \pb_0^2 \left(\sqrt{1
    -{\tau_0^2+4\,\pb_0^2\tau_0-8\,\delta_H^2\over2\pb_0^2}}-1\right),
    \label{ellipse}
\eeq
and satisfy the single pair of additional inequalities (\ref{deltav}).

We now need to find out whether $\tau$ and $\tau_0$ can be chosen
to meet these requirements. Clearly, not all points in the region
(\ref{ellipse}) satisfy (\ref{deltav}). This can be seen setting
$\tau_0 = 0$, in which case we get $\half\,\tau^2 + \pb_\pm^2\tau
< 0$, which has no positive solutions for $\tau$; in fact, if we
consider (\ref{deltav}) as a second-order inequality for $\tau$
for a given (positive) value of $\tau_0$,
\beq
    (\bb_0^2\tau_0+\half)\,\tau^2 + [\bb_\pm^2\tau_0^2
    + (1-2\,\delta_v^2)\,\tau_0+\pb_\pm^2]\,\tau
    + (\half\,\tau_0^2+\pb_0^2\tau_0) < 0\;, \label{newform}
\eeq
we see that there are positive solutions for $\tau$ only if
the coefficient of the linear term is a sufficiently large
negative number. We can also see directly from (\ref{ineq2}) that
we must have $\delta_v^2 \gg 1$. For any positive $\tau_0$, as
$\delta_v$ becomes very large, the range of values of $\tau$ which
satisfy (\ref{newform}) becomes the full positive half-line
$(0,\infty)$, so all we need to do is pick a large enough
$\delta_v$ for all conditions to be satisfied in a region of the
$\tau$-$\tau_0$ plane. The fact that $\delta_v$ must be large for
suitable states to exist is an example of the uncertainty
relations putting bounds on products of tolerances, and a
consequence of the fact that we are requiring $\delta_p$ to be
small.

We finish this section by giving a set of parameter values (different
from the ones in Fig.~\ref{f1}) which satisfy the relationships
mentioned above, and allow us to choose smaller values for $\tau$
and $\tau_0$, for use in later parts of the paper. If for
simplicity we set $\pb_+ = \pb_-$ and $\bb_+ = \bb_-$ (the two
inequalities (\ref{newform}) or their equivalents become a single
one), it is natural to make the choices
\beq
    \pb_\pm=1\;,\quad \pb_0=\sqrt2\;,\quad \bb_\pm=1\;,\quad \bb_0=1
    \label{par1}
\eeq
for the classical phase space point (notice that all inequalities are
invariant under an appropriate rescaling of the variables involved;
this can be used for example to assign any desired value to $\pb_0$,
which we think of as an arbitrary `global scale'), and
\beq
    \delta_p=0.15\;,\quad \delta_v=12\;,\quad \eps_H=\delta_H=0.25
    \label{par2}
\eeq
for the relevant tolerances. Then, a pair of width parameters that
solves the inequalities is
\beq
    \tau = \tau_0 = 0.025\;. \label{par3}
\eeq

\section{Group-Averaged Coherent States}
\label{sec:5}

\noindent In the Dirac approach to the quantization of first-class
constrained systems, a {\it physical state} $\Psi_{\rm phy}$ is
one which is annihilated by the Hamiltonian constraint,
$\hat{H}\cdot\Psi_{\rm phy} = 0$. It is a general feature of
constrained systems that physical states will {\it not} belong to
the kinematical Hilbert space ${\cal H}_{\rm kin}$, and the
Refined Algebraic Quantization program has been developed to deal
with such instances \cite{Marolf}. The general strategy is the
following. One first identifies a dense subspace $\Phi$ of the
Hilbert space ${\cal H}_{\rm kin}$, where the Hamiltonian is well
defined (that is, $\Phi$ is contained in the domain of $\hat H$).
The physical states will then belong to the dual $\Phi^*$, which
satisfies the inclusion relation $\Phi \subset {\cal H}_{\rm kin}
\subset \Phi^*$. The choice of the dense subset $\Phi$ is a
delicate matter, and normally has some physical implications
\cite{Marolf}, but for our purposes we are only interested in
looking at coherent states which have some ``nice'' properties
that make them well behaved, and those states belong to most
choices of the space $\Phi$. The procedure we will follow for
implementing the Dirac requirement that physical quantum states be
annihilated by the constraint operators is the so-called {\it
group averaging}, a particularly popular approach within the
general RAQ program.

The intuitive idea is to start with a state $\psi\in \Phi$ and
average it over the group generated by the constraint. If all goes
well, one will end up with an element of $\Phi^*$ satisfying the
constraint. Our strategy will be very naive. Starting with the
kinematical coherent state (\ref{psikin}), group averaging gives
us the physical coherent state $\psi_{\rm phy} = N^{-1}
\int\dd\lambda\, \ee^{-\ii\lambda\hat H}\, \psi_{\rm kin}$, where
$N$ is a normalization factor. Since in $p_i$-space the constraint
is a multiplication operator, the integration is straightforward
and gives a delta function $\delta(p_0^2-p_+^2-p_-^2)$, so we can
identify $\psi_{\rm phy}(p_\pm)$ with $\psi_{\rm
kin}((p_+^2+p_-^2)^{1/2},p_\pm)$, up to a new normalization factor
to be determined,\footnote{The group average procedure, in a
strict sense, will yield also a contribution from the past null
cone, but since we are considering Gaussian wave functions that
are almost zero there, the contribution from that null cone is
negligible.}
\bea
    \psi\phy(p_\pm)
    = & &{\cal N}\, \ee^{
    - [(p_+^2+p_-^2)^{1/2}-\pb_0]^2/2\tau_0
    - (p_+ - \pb_+)^2/2\tau_+ -(p_- - \pb_-)^2/2\tau_- } \times
    \nonumber\\
    & &\times\ \ee^{ -\ii\,[(p_+^2+p_-^2)^{1/2} - \pb_0]\,\bb_0
    - \ii\,(p_+ - \pb_+)\,\bb_+ - \ii\,(p_- - \pb_-)\,\bb_- }\;.
    \label{psiphy}
\eea
Of course, strictly speaking what one is doing is defining a function on
the future null cone ($p_0^2-p_+^2-p_-^2 = 0$, $p_0>0$), with
coordinates $p_\pm$. The inner product induced by the group averaging
procedure is then of the form
\beq
    (\psi|\phi\rangle =
    \int \frac{\dd p_+\,\dd p_-}{2\,(p_+^2+p_-^2)^{1/2}}\;
    \psi\phy^*(p_\pm)\,\phi\phy(p_\pm)\;.
    \label{inner}
\eeq
Again, we want to calculate the expectation values and
fluctuations of the physical observables $\hat p_\pm$ and $\hat
v_i$, and compare them to the values computed with the kinematical
coherent states. The strategy here will be to represent the Dirac
observables in the physical Hilbert space making use of the
measure (\ref{inner}) induced by the group average and the general
form of the operators given by (\ref{operators}). The momentum
observables act as multiplication operators; to find the action of
the configuration-like ones we use again (\ref{operators}), where
now $\mu(p_\pm) = \half\, (p_+^2 + p_-^2)^{-1/2}$ is the measure
appearing in (\ref{inner}). This gives, on physical states,
\beq
    \hat v_\pm\, \psi(p_\pm) = \ii \sqrt{p_+^2+p_-^2}
    \,\partial_\pm \psi(p_\pm)\;,\qquad
    \hat v_0\, \psi(p_\pm) = \ii\,
    (p_+\partial_- - p_-\partial_+)\, \psi(p_\pm)\;. \label{33}
\eeq
The first step, before calculating expectation values and fluctuations,
is to calculate $\cal N$. However, because of the presence of the
expression $(p_+^2+p_-^2)^{1/2}$ in the exponent of $\psi\phy$, as well as
in the measure, the relevant integrals cannot be calculated exactly, and
we need to use an approximation.

If the state $\psi\phy$ is narrowly peaked in the $p_\pm$ directions, we
can make use of the assumption $\tau \ll \pb_0^2$ in the approximation,
and keep only the first few terms in the expansion of every result in
powers of $\tau/\pb_0^2$. To see how to implement this idea, let us
parametrize the deviation of $p_\pm$ from the classical values $\pb_\pm$
by a new pair of variables,
\beq
    x:= {p_+\over\pb_0}\cos\bar\theta
    + {p_-\over\pb_0}\sin\bar\theta - 1\;, \qquad
    y:= -{p_+\over\pb_0}\sin\bar\theta
    + {p_-\over\pb_0}\cos\bar\theta\;,
\eeq
where $\bar\theta$ is the angle between the vector
$(\pb_+,\pb_-)$ and the positive $p_+$-axis, and the origin in the
$(x,y)$ coordinates corresponds to the point where the Gaussian is
peaked. We can,  equivalently solve
\beq
    p_+ = (x+1)\,\pb_+ - y\,\pb_-\;,\qquad
    p_- = (x+1)\,\pb_- + y\,\pb_+\;. \label{35}
\eeq
Intuitively, $x$ and $y$ are the deviations from $\pb_\pm$ along the
radial direction in the $p_\pm$ plane and along the $p_+^2 + p_-^2 =
\pb_0^2$ circle, respectively, both expressed in units of $\pb_0$. Using
these definitions we get, e.g., that $p_+^2 + p_-^2 = \pb_0^2\,
[(1+x)^2+y^2]$, and in terms of the variables $x$ and $y$ the physical
state can be written as
\bea
    \psi\phy(x,y)
    = & &{\cal N}\, \ee^{-\{[(1+x)^2+y^2]^{1/2}-1\}^2\pb_0^2/2\tau_0
    - (x^2+y^2)\pb_0^2/2\tau } \times \nonumber\\
    & &\times\ \ee^{-\ii\,\{[(1+x)^2+y^2]^{1/2} - 1\}\,\pb_0\bb_0
    - \ii\,(\pb_+x - \pb_-y)\,\bb_+ - \ii\,(\pb_-x + \pb_+y)\,\bb_-}\;;
    \label{psiphyxy}
\eea
in the remainder of this paper, we will assume for simplicity that
$\tau_+ = \tau_- =: \tau$. Notice also that we will always assume that
the parameters satisfy the classical constraint $\pb_0^2=\pb_+^2+\pb_-^2$.

Our strategy for obtaining expressions for $\cal N$, as well as all
expectation values and fluctuations, will be the following. All
calculations can be reduced to integrals of the form
\beq
    \int {\dd x\,\dd y\over[(1+x)^2+y^2]^{1/2}}
    \,\psi(x,y)^*\,\hat O\,\psi(x,y)\;,
\eeq
where $\hat O$ may be a combination of multiplication and derivative
operators. Since $\psi^*\hat O\psi$ always contains the Gaussian factor
$\exp\{- (x^2+y^2)\pb_0^2/\tau\}$ which is peaked at small values of $x$
and $y$, we will expand all remaining factors in the integrand, including
the measure, in powers of $x$ and $y$, and calculate the integral term by
term in the expansion. Each non-vanishing term is a Gaussian integral of
the form
\beq
    \int_{-\infty}^{+\infty} \dd x\,x^{2n}\,\ee^{-x^2\pb_0^2/\tau}
    \int_{-\infty}^{+\infty} \dd y\,y^{2m}\,\ee^{-y^2\pb_0^2/\tau}
    = {\pi\,(2n-1)!!\,(2m-1)!!\over2^{n+m}}
    \left({\tau\over\pb_0^2}\right)^{\!n+m+1}, \label{Gaussian}
\eeq
with a coefficient that depends on the parameters in $\psi$
and, most importantly, contains one inverse power of $\tau$ for
each derivative in $\hat O$; if we want to keep terms up to a
certain order in $\tau/\pb_0^2$ when adding the integrals of the
type (\ref{Gaussian}), we need to take this into account. With the
exception of $(\Delta\hat H)\kin^2$ and $(\Delta\hat v_i)\kin^2$,
the expectation values and fluctuations of our operators in the
kinematical states contain a single power of $\tau$; therefore, in
order to compare those results with the corresponding ones for
physical states, for the latter in practice we need to keep terms
in our expansions up to the two leading orders in $\tau/\pb_0^2$.

The normalization factor $\cal N$ is the simplest expression to
calculate. Since
\beq
    {\cal N}^{-2} = {\pb_0\over2}
    \int {\dd x\,\dd y\over[(1+x)^2+y^2]^{1/2}}
    \,|\psi\phy(x,y)|^2
    = {\pi\tau\over2\pb_0} \left[ 1
    - \left( {\pb_0^2\over\tau_0} - {1\over2} \right)
    {\tau\over2\pb_0^2} + {\cal O}(\tau^2) \right],
\eeq
the result is
\beq
    {\cal N}^2 = {2\pb_0\over\pi\tau} \left[ 1 + \left(
    {\pb_0^2\over\tau_0}-{1\over2}\right) {\tau\over2\pb_0^2} \right]
    + {\cal O}(\tau)\;. \label{normal}
\eeq
With $\cal N$, we can now calculate other expectation values. It is
useful to first calculate the auxiliary quantities
\bea
    & &\ev{x} = -{\tau\over2\pb_0^2}
    + \left({\pb_0^2\over\tau_0}-1\right) {\tau^2\over4\pb_0^4}
    + {\cal O}(\tau^3)\;,
    \qquad\quad \ev{y} = 0\;, \nonumber\\
    & &\ev{x^2} = {\tau\over2\pb_0^2}
    - \left({\pb_0^2\over\tau_0}-1\right) {\tau^2\over2\pb_0^4}
    + {\cal O}(\tau^3)\;,
    \qquad\quad \ev{y^2} = {\tau\over2\pb_0^2} - {\tau^2\over4\pb_0^4}
    + {\cal O}(\tau^3)\;,
\eea
with which we find for the $\hat p_\pm$,
\bea
    \ev{\hat p_\pm}\phy &=&
    \int {\dd p_+\dd p_-\over2\,(p_+^2+p_-^2)^{1/2}}\,
    p_\pm\,|\psi\phy|^2 \nonumber\\
    & = &\pb_\pm\,(\ev{\hat x} + 1)
    = \pb_\pm \left[ 1 - {\tau\over2\pb_0^2}
    + \left({\pb_0^2\over\tau_0}-1 \right)
    {\tau^2\over4\pb_0^4} + {\cal O}(\tau^3) \right],\label{evp}
\eea
and for the fluctuations of the same operators,
\bea
    \ev{\hat p_\pm^2}\phy
    & = & \pb_\pm^2\,(1 + 2\ev{\hat x} + \ev{\hat x^2})
    + \pb_\mp^2\,\ev{\hat y^2} \vphantom\int \nonumber \\
    & = & \pb_\pm^2 + {\pb_\mp^2 - \pb_\pm^2\over2\pb_0^2}\, \tau
    - {\pb_\mp^2\,\tau^2\over4\pb_0^4} +{\cal O}(\tau^3) \nonumber \\
    (\Delta\hat p_\pm)^2\phy
    & = & \ev{\hat p_\pm^2}\phy - \ev{\hat p_\pm}\phy^2
    = {\tau\over2}
    + \left[{\pb_\pm^2\over\pb_0^2}
    - \left({1\over2}+{\pb_\pm^2\over\tau_0}\right) \right]
    {\tau^2\over2\pb_0^2} + {\cal O}(\tau^3)\;.
\eea
As for the Hamiltonian constraint, the exact physical state $\psi\phy$
of course gives $\ev{\hat H}\phy = 0$ and $(\Delta\hat H)\phy^2 = 0$, by
definition.

The $\hat v_i$ operators contain derivatives with respect to $p_\pm$;
to calculate their expectation values it is useful to recast those in
terms of $(x,y)$, and we find, from (\ref{33}) and (\ref{35}),
\bea
    \hat v_\pm\psi(x,y) &=& \ii\, \sqrt{(1+x)^2+y^2}
    \left({\pb_\pm\over\pb_0}\,\partial_x\mp {\pb_\mp\over\pb_0}
    \partial_y\right) \psi \nonumber \\
    \hat v_0 \psi(x,y) &=& \ii\,
    \big[ -y\, \partial_x + (1+x)\, \partial_y \big]\,\psi\;.
\eea
Then for the expectation values we find
\bea
    \ev{\hat v_\pm}\phy
    & = & {\ii\over2} \int \dd x\,\dd y\,\psi\phy^*
    \left({\pb_\pm\over\pb_0}\,\partial_x\mp {\pb_\mp\over\pb_0}
    \partial_y\right) \psi \nonumber\\
    & = & (\pb_0 \bb_\pm + \pb_\pm \bb_0)
    - (\pb_0 \bb_\pm + 2\,\pb_\pm \bb_0)\, {\tau\over4\pb_0^2}
    + {\cal O}(\tau^2)\;, \nonumber\\
    \ev{\hat v_0}\phy
    & = & {\ii\over2}
    \int {\dd x\,\dd y\over[(1+x)^2+y^2]^{1/2}}\,\psi\phy^*
    \Big(-y\,\partial_x + (1+x)\,\partial_y\Big) \psi\phy \nonumber\\
    & = & (\pb_+\bb_- - \pb_-\bb_+) \left(1-{\tau\over2\pb_0^2}
    + {\cal O}(\tau^2)\right); \label{evv}
\eea
similarly, the fluctuations of these operators are, respectively,
\bea
    & &(\Delta\hat v_\pm)\phy^2
    = {\pb_0^2\over2\tau} + {1\over4}
    \left(1 - {\pb_\pm^2\over\pb_0^2} + {2\pb_\pm^2\over\tau_0}\right)
    + {\cal O}(\tau)\;, \nonumber\\
    & &(\Delta\hat v_0)\phy^2
    = {\pb_0^2\over2\tau} - {1\over4} + {\cal O}(\tau)\;.
\eea

We now have all the results we need to write down the conditions
that the physical states be semiclassical, but we can also bypass
this step by a direct comparison between the expectation values
for the $O_\alpha$ in the two sets of states, \bea
    & &\ev{\hat p_\pm}\phy - \pb_\pm
    = -{\tau\over2\pb_0^2} + {\cal O}(\tau^2)\\
    & &\ev{\hat v_\pm}\phy - \bar v_\pm
    = -(\pb_0\bb_\pm^2+2\,\pb_\pm\bb_0^2)\,{\tau\over4\pb_0^2}
    + {\cal O}(\tau^2)\;,
\eea
and between the corresponding fluctuations,
\bea
    & &(\Delta\hat p_\pm)\phy^2 - (\Delta\hat p_\pm)\kin^2
    =  -\left( {\pb_\pm^2\over\tau_0} + {1\over2}
    - {\pb_\pm^2\over\pb_0^2} \right) {\tau^2\over\pb_0^2}
    + {\cal O}(\tau^3) \label{reldelp}\\
    & &(\Delta\hat v_\pm)\phy^2 - (\Delta\hat v_\pm)\kin^2
    = -{1\over4} \left(1+{\tau_0\over\tau}+{\tau\over\tau_0}\right)
    - {\pb_\pm^2\over4\,\pb_0^2} - \bb_\pm^2\tau_0 + {\cal O}(\tau)\;.
     \label{reldelv}
\eea
The first pair of relationships tell us that the physical
coherent states have expectation values for the $O_\alpha$ that
are close (to order $\tau$) to the classical values; that fact,
together with the observation that the right-hand sides of the
last two equations are negative (provided that $\tau_0 <
2\,\pb_0^2$), shows that the physical states are also
semiclassical, with respect to the same tolerances used for the
kinematical state and values for $\eps_p$ and $\eps_v$ that can be
chosen to be small of order $\tau$. In particular, if we choose as
reference classical state the one specified by (\ref{par1}), and
our criteria for a semiclassical state include the choices
(\ref{par2}) plus, for example,
\beq
    \eps_p = 0.01\;, \quad \eps_v = 0.02\;,
\eeq
then the state $\psi\phy$ constructed using the width parameters
given in (\ref{par3}) is semiclassical.

Thus, with this choice of widths, the kinematical state is
compatible with the physical one in the less restrictive of the
two possible definitions mentioned in Sec \ref{sec:3}. In the more
restrictive sense, however, it is in general not a good
approximation, because the right-hand side of (\ref{reldelp}) is
not necessarily small compared to $(\Delta\hat p_\pm)\phy^2$, nor
that of (\ref{reldelv}) compared to $(\Delta\hat v_\pm)\phy^2$.
For the former to happen, the widths would have to satisfy $\tau
\ll \tau_0$ (or $\pb_\pm$ would have to be very small, but this
cannot be true for both signs), while for the latter to happen,
they would have to satisfy (either $\tau \approx \tau_0$ or) $\tau
\ll \tau_0 \ll \pb_0^2$.

\section{Time Evolution}
\label{sec:6}

\noindent In the previous sections we have considered two kinds of
coherent states, kinematical and ``group averaged" ones. The first
kind are coherent states defined on the full phase space and
peaked around a point of the constraint surface. They do not
satisfy the quantum constraint. The second kind of states are
constructed out of the first kind by averaging over the orbit
generated by the constraint. Even though the second kind of states
are in a sense {\it dynamical} since one can think of the
constraint (for totally constrained systems, such as this one) as
generating time evolution, in a strict sense they are `frozen'.
Once the quantum state belongs to the physical Hilbert space, the
notion of `gauge' time evolution is lost. There is no notion of
time evolution, no dynamics. Within this perspective, the
observables that we have considered, namely Dirac observables, are
constant along the orbits of the constraint and are therefore
constants of the motion according to the same dynamical
interpretation. This is the well known ``frozen dynamics
formalism" for time-reparametrization invariant constrained
systems.

Is there a way of recovering dynamics from such a formalism? Again
the answer is in the affirmative and several solutions have been
known for a while \cite{rovelli,ATU}. We shall consider one of
those possibilities. For this we shall proceed in two steps. In
the first part of the remainder of this section we recast the
quantum constraint equation in such a way that it resembles a
Schr\"odinger equation. In the second part we consider observables
(different from the ones already considered) that capture the
intuitive notion of time evolution and that will allow us to ask
dynamically meaningful questions such as: given a coherent state
with a small spread in the anisotropies at, say, the Plank time,
we would like to know how much time later (as measured by a proper
measure of the total volume) the state will spread. Based on the
previous results obtained so far, where we found a range of
parameters that produce acceptable coherent states, we shall try
to give answers to these type of questions.

\subsection{Schr\"odinger Equation}

\noindent The idea here is to consider states at the kinematical
level that are in the $p$ representation for the $\pm$ sector but
are functions of the spatial volume and will therefore be diagonal
on $\beta_0$. That is, we will consider functions of the form
$\Psi(\beta_0,p_\pm)$. We want to consider the quantum constraint
$\hat{H} = -\half\,(\hat{p}_0^2+\hat{p}_+^2+\hat{p}_-^2) = 0$,
rewritten as
\beq
    \hat{H} = -\half\,(\hat{p}_0+\sqrt{\hat{p}_+^2+\hat{p}_-^2})
    (\hat{p}_0-\sqrt{\hat{p}_+^2+\hat{p}_-^2})\;,
\eeq
and look for solutions to the quantum constraint that are of
the form
\beq
    (\hat{p}_0-\sqrt{\hat{p}_+^2+\hat{p}_-^2})\cdot\Psi(\beta_0,p_\pm)
    =0\;. \label{sqrt}
\eeq
The choice of relative sign in (\ref{sqrt}) is equivalent to
the choice we made earlier of the future (as opposed to past)
light cone in $p$-space, in the context of group-averaged coherent
states. In a sense we are following the original ``already
parametrized" program of ADM.

Note that this strategy is equivalent to considering the reduced
phase space, described by true Dirac observables, and defining
wave functions of them that have a ``time dependence" as they are
also functions of $\beta_0$. Among these states, we can look for
coherent states peaked around `physical' phase space points. We
start therefore with the reduced phase space $\Gamma_{\rm r}$ for
which we will use the coordinates $\Gamma_{\rm r} =
\{(p_I,\beta_I),\ I = \pm\}$, and the Hilbert space ${\cal H}_{\rm
r} = {\rm L}^2(\real^2,\dd p_+\dd p_-)$. The sector of the solutions
to the Hamiltonian constraint now takes the form of a
Schr\"odinger equation,
\beq
    \ii\,{\partial\over\partial\beta_0}\,\Psi(\beta_0,p_\pm)
    = \sqrt{p_+^2+p_-^2}\,\Psi(\beta_0,p_\pm)\;. \label{timev}
\eeq
That is, any (normalizable) function $\psi(p_\pm)$ gives rise
to a solution to the quantum constraint via the ``time evolution
equation" (\ref{timev}), whose solution can easily be written as
\beq
    \Psi(\beta_0,p_\pm)
    =e^{-\ii\sqrt{p_+^2+p_-^2}\;\beta_0}\psi(p_\pm)\;.
\eeq

The strategy now is to start with a coherent state ``intrinsic" to
the $p_\pm$ plane (as a coordinatization of the reduced
configuration space) and then construct the corresponding physical
state. This choice is motivated by the fact that the quantities
$\beta_\pm$ have a clear space-time interpretation as the
anisotropies  of the cosmological model. Let us then consider, as
initial state,
\beq
    \psi_0(p_\pm) = \prod_{i=\pm}\,{\cal N}_i
    \,\exp[-(p_i-\bar{p}_i)^2/2\tau_i - \ii\, (p_i -
    \bar{p}_i) \bar\beta_i]\;,
\eeq
which is a coherent state peaked around the point $(\bar{p}_\pm,
\bar\beta_\pm)$ on the reduced phase space, and with spread dictated
by the parameters $\tau_i$. We get then
\bea
    \Psi_0(\beta_0,p_\pm) &=&
    \ee^{-\ii\sqrt{p_+^2+p_-^2}\,\beta_0}\psi_0(p_\pm)\nonumber\\
    &=& \prod_{i=\pm}\,{\cal N}_i
    \,\exp\left[-(p_i-\bar{p}_i)^2/2\tau_i - \ii\, (p_i -
    \bar{p}_i) \bar\beta_i \right]
    \ee^{-\ii \sqrt{p_+^2+p_-^2}\,\beta_0}.
\eea
Let us note that the general solution should be a function of
$(\beta_0-t_0)$, for $t_0$ the ``initial value" of the parameter
$\beta_0$. Now, when $\beta_0 = 0$, the {\it classical} volume
element is given by $\sqrt{\det q}=1$ in Planck units. Thus, the
``Planck epoch" is given by the values of $\beta_0 = 0$ (recall
that the ``singularity" is at $\beta_0 \to -\infty$). We shall
then, in the remainder of this section, take $t_0 = 0$, and consider
the ``initial wave function" at $t_0 = 0$ as being defined at
the Planck epoch.

\subsection{Time Evolution of Observables}

The next step is to consider observables. On the one hand we know
from the general theory of constrained systems that physical
observables are those operators that preserve the space of
physical states. Properly represented Dirac observables have that
property and that is why we have considered them in previous
sections. On the other hand, we are interested in disentangling
dynamical information from the particular (Schr\"odinger-like)
representation of this part. This means that we need observables
that capture the notion of time evolution. Such observables exist
and are sometimes referred to as {\it evolving constants of the
motion} \cite{rovelli}. The physical states considered in this
part are solutions to the equation (\ref{timev}) so we have to
define observables that preserve the space of such solutions. If
we were to consider the operator $\hat\beta_0$ (that naturally
acts as a multiplication operator), we would be thrown out of the
physical Hilbert space. It is not a physical observable. What we
need to do is to consider the following situation: let us assume
that the phase space function $\beta_0$ is an (internal) time
parameter, and that we can consider a new kind of observables
`defined at time' $\beta_0 = t$. That is, we fix a $\beta_0 = t$
slice in the configuration space $(p_\pm,\beta_0)$ and consider
the restriction of the wave function to that slice,
$\Psi(\beta_0=t,p_\pm)$. We can then act with an operator (that
might be `$\beta_0$-dependent') and consider the time evolution of
the resulting wave function, thus obtaining again a physical
state. This is nothing but the ordinary prescription for defining
Heisenberg observables in ordinary quantum mechanics.

Consider an observable that depends explicitly on $\beta_0$ and
does not commute with the constraint (for instance the volume form
$V(\beta_0) = \ee^{3\beta_0}$). The strategy is to define a
one-parameter family of observables $V(t) = V(\beta_0=t)$ as
Heisenberg observables and then consider the corresponding
operators. Note that $\hat{V}(t_1)\neq\hat{V}(t_2)$ if $t_1\neq
t_2$, as physical operators on the physical Hilbert space. The
observables that we will be interested in are of two kinds. On the
one hand, we shall consider the `configuration observables'
$\hat{p}_\pm$ that are constants of the motion since they commute
with $\hat{H}$, and on the other hand those operators that measure
the anisotropies, namely the operators $\hat\beta_\pm$. These are
not explicitly `time dependent', but we expect both expectation
values and dispersions to be changing quantities.\footnote{Let us
qualify the statement: We expect that the one-parameter family of
numbers $\langle \hat\beta_\pm \rangle_t$, corresponding to the
one-parameter family of operators $\hat\beta_{\pm,t}$, will depend
on the value of $t$.}

Let us then compute the expectation values and quantum
fluctuations for the operators $\hat\beta_\pm = \ii\,\partial/
\partial p_\pm$. The expectation value will take the form
\beq
    \langle \hat\beta_+\rangle_t = {\cal N}^2\left[\int \dd^2p_\pm\;
    \ee^{-(p_+-\bar{p}_+)^2/\tau}\,\frac{p_+}
    {\sqrt{p_+^2+p_-^2}}\right] t + \bar\beta_+\;.
\eeq
Note that there will be a shift of the `peak' of the wave
function, since the expectation value will have a linear $t$
dependence. This is to be expected since classically we know that
the system evolves in a way that the $\beta_\pm$ depend linearly
on time. In order to compute the coefficient we need to make some
approximations as in previous sections. Making use of the same
assumptions, we arrive for the integral to the following
expression,
\beq
    \langle \hat\beta_+ \rangle_t = \bb_+ + t\,{\cal N}^2
    \pb_0^2\int\dd x\,\dd y\;
    \ee^{-\pb_0^2(x^2+y^2)/\tau}\;\frac{(1+x)\bar{p}_+-y
    \bar{p}_-}{\pb_0\,[(1+x)^2+y^2]^{1/2}}\;.
\eeq
This can be approximated, as a power series expansion in $\tau$,
by the expression
\beq
    \langle \hat\beta_\pm \rangle_t = \bb_\pm
    +\frac{\bar{p}_\pm}{\pb_0}\,\left[1-\frac{1}{4}\,
    \frac{\tau}{\pb_0^2}-\frac{3}{8}\,\frac{\tau^2}{\pb_0^4}+{\cal
    O}(\tau^3)\right] t\;.
\eeq
Note that at zeroth order in $\tau$ the speed of the peak
coincides precisely with the classical value. Let us now compute
the fluctuations of the operators and consider the expectation
value $\langle \hat\beta^2_+\rangle$, to second order in $\tau$
as,
\beq
    \langle \hat\beta^2_+\rangle_t = \frac{1}{2\tau}+\bar\beta^2_+ +
    2 \bar\beta_+\;\frac{\bar{p}_+}{\pb_0} \left[1-\frac{1}{4}\,
    \frac{\tau}{\pb_0^2}-\frac{3}{8}\,\frac{\tau^2}{\pb_0^4}\right] t
    +\frac{\bar{p}^2_+}{\pb_0^2}\left[1 -\frac{1}{2}\,
    \frac{\tau}{\pb_0^2}+\frac{33}{4}\,
    \frac{\tau^2}{\pb_0^4}\right] t^2\;.
\eeq
From here we can then compute the fluctuations of the anisotropy
observables. We have then,
\beq
    (\Delta\beta_\pm)_t^2 = \langle\hat{\beta}_\pm^2\rangle_t-
    \langle\hat{\beta}_\pm\rangle^2_t=\frac{1}{2\tau}+\frac{143}{16}\,
    \frac{\bar{p}_\pm^2}{\bar{p}_0^6}\,\tau^2\,t^2\;,
\eeq
where we keep terms that are at most quadratic in $\tau$. We
see that, as in the case of a non-relativistic free particle, the
spread of the Gaussian grows quadratically with `time' $t$. The
next step is to estimate the value $t_{\rm max}$ of the maximum
allowed value for $t$ before the wave packet spreads considerably.
For that we use the values of the parameters $\bar{p}_\pm$,
$\bar{p}_0$ and $\tau$ chosen in (\ref{par1}) and (\ref{par3}).
The reason for this is
that we expect that the conditions imposed on the expectation
values and fluctuations of the relevant operators yield realistic
values for the parameters defining the coherent states; even though
in this part we start with intrinsic coherent states at `Planck
time', the resulting wave function is closely related to the
physical coherent states of Sec.~\ref{sec:5}, and it seems
reasonable to employ the parameters used there. If we define the
time for the wave packet to spread significantly as
the value for $t_{\rm max}$ such that
$(\Delta\beta_\pm)_{t_{\rm max}}^2$ is about twice as large as
$(\Delta\beta_\pm)_{t_{0}}^2\approx 20$, with the values of
the parameters given in Sec~\ref{sec:4} it is easy to see that
$t_{\rm max}\approx 170$. That is, the time for which the packet
spreads to about twice its size is within 200 times the
fundamental time unit, the Planck time. With a different
choice of parameters that yield smaller values of $\tau$, one
might have slightly longer spread times, but within the same order
of magnitude. Another possibility would be to take the values used
in Fig.~\ref{f1} and choose $\tau = .03$, which is near the high
end of the range of allowed values. In this case the time
$t_{\rm max}\approx 65$, again of the same order
of magnitude as with the previous choice.

One might ask about the possibility of having semiclassical states
that approximate the $\beta$'s more accurately, which means
smaller initial fluctuations (of the order of $1/\tau$). This
would require large values of $\tau$, for which the expansion in
powers of $\tau$ would not be justified. In that case, one would
need to expand expectation values and fluctuations of observables
in inverse powers of $\tau$, and use those in the set of
inequalities that express the semi-classicality conditions. With
this choice, one might be able to construct states that take
longer to spread. We shall not explore this possibility further.
Let us end this section with a remark. Even though we have found a
`time' (as measured in terms of the volume element) for which the
initial wave-packet spreads, we would not like to suggest that
this has a direct significance for the spread of the wave-function
in realistic cosmological models, since we have neglected any
matter and the effect that this may have in possible decoherence
effects on the wave function.

\section{Conclusion and Outlook}
\label{sec:7}

\noindent In this paper we have only considered the simplest type
of anisotropic cosmological models, the vacuum ones of Bianchi type I.
For our purposes, the simplicity of these models lies in the fact that
their phase space is a vector space, their only constraint is
quadratic, and it depends on the $p$ variables only; in this
sense, these models are similar to the general class of models
treated in Ref \cite{abc}, in which kinematical and physical
coherent states are also discussed. However, on the one hand our
Hamiltonian and the set of physical observables we chose are not
of the type used there, and on the other hand the fact that the
Hamiltonian itself is the constraint raises the usual issue of
time evolution in generally covariant systems, and led us to the
construction of a third type of state that we called ``intrinsic",
with which physical predictions on the evolution of physical
observables can be attempted. The relevance of our findings in
this simple model for the ultimate goal of constructing realistic
semi-classical states in full quantum gravity still needs to be
explored. However, we hope that these first steps might shed some
light on the program.

To explore further the relationships between kinematical and
physical coherent states as candidate semiclassical states for
gravity models, a natural extension of this work would be to
consider other Bianchi type models. As soon as one starts doing
that, one is faced with the fact that the potential $U(\beta^i)$
does not vanish, which implies that the quantum Hamiltonian is
no longer a multiplication operator. Fortunately, in some cases
this difficulty can be circumvented. As is well known, the
dynamics of a system on a configuration space with a given metric
subject to a potential $U$, can be mapped into that of a system
with a new metric (the Jacobi metric) and vanishing potential.
In our case, the transformation of a Hamiltonian like (\ref{H})
into an equivalent one describing a free particle follows the
standard procedure, but what one usually ends up with is a curved
Jacobi metric replacing $\eta_{ij}$. What is not so obvious is the
fact that in the case of many homogeneous cosmologies, known as
{\it diagonal, intrinsically multiply transitive\/} (DIMT) models,
the new metric is still flat \cite{ATU}. These models include the
following cases: Bianchi types I and II; sub-families of Bianchi
types III, VIII and IX defined by $\beta^- = 0$ (the Taub model);
Kantowski-Sachs spacetimes; and Bianchi type V models with
$\beta^+ = 0$. For Bianchi type I and II, the minisuperspaces are
3-dimensional, parametrized by $\beta^0$, $\beta^+$ and $\beta^-$
in the Misner scheme; for the remaining DIMT models they are
2-dimensional, parametrized by an appropriate subset of the
$\beta^i$ \cite{Bianchi}. Note that even though the type III
and V models are class B, which means that in general the
Hamiltonian description is more subtle \cite{ryan}, these
particular cases do not suffer from the problems described
in Refs \cite{torre1,ryan}.

Thus, DIMT models are natural candidates for a first extension of
our work. When one carries out the transformation to a new set
of phase space variables $(\tb^i, \tp_i)$ in which the potential
term in (\ref{H}) vanishes and the metric again has the form
$\eta_{ij}$, the difference between the various DIMT models shows
up in the fact that the allowed ranges of the $\tp_i$ are not
the whole real line \cite{ATU}. Therefore, those variables cannot
be regarded as momenta in the regular sense of elements of the
cotangent space, and in the quantum theory the corresponding
operators in the $\tb_i$-representation cannot be the usual
derivative operators. (Note: this feature, which also occurs in
full quantum gravity, has been addressed in that context by
Klauder \cite{klauder1} and Loll \cite{loll}.) We have then one
more reason to set up the quantum theory in the $\tp$-representation
\cite{ATU}. This time, however, the ``configuration space" for
these models is a proper subset of 2+1 Minkowski space, with
restrictions on the values of the $\tp_i$ depending on the model
under consideration, and the constraint surface is the intersection
of the (future) light cone of the origin with this subset.

With this setup, one can define suitable generalizations of
coherent states in the configuration spaces of the various DIMT
models, which now are non-trivial spaces (for example, a half-line
rather than the whole real line), and analyze their properties as
candidate semiclassical states. One possible strategy is to construct
such states as suggested by Klauder's work \cite{klauder2}, and
search for a proper subset of states that are consistent with our
physical requirements, but we shall leave that investigation for
a future publication. As a further step, one can envision extending
the work to models which cannot be formulated as free particles on
portions of Minkowski space, which will require new techniques to
handle the quantum constraint (it would be interesting, for example,
to study the connection with Kiefer's use of the ``principle of
constructive interference" to build wave packets in minisuperspace
\cite{kiefer2} and other models for the emergence of classical
Friedmann-Robertson-Walker spacetimes \cite{kim}), and then
possibly some non-homogeneous, midisuperspace models. In terms of
our original goal of understanding the relationship between the
kinematical and physical semiclassical states for quantum
gravity, the hope is that at some point one may see a pattern
that will allow us to make statements of a more general nature.

\section*{Acknowledgements}

We are grateful to Abhay Ashtekar for many discussions, which
provided the motivation for this work, and for suggestions. We
also thank Michael Ryan for comments and David Sanders for help
with the figure. Partial support for this work was provided by
NSF grant PHY-0010061, CONACyT grant J32754-E and DGAPA-UNAM
grant 112401.



\begin{thebibliography}{999}
\raggedright

\bibitem{torre1} Fels M E and Torre C G 2002
``The principle of symmetric criticality in general relativity,''
{\sl Class.\ Quantum Grav.}\ {\bf19} 641,
{\tt arXiv:gr-qc/0108033}

\bibitem{mike} Kucha\v{r} K V and Ryan M P 1989
``Is minisuperspace quantization valid? Taub in Mixmaster,'' {\sl
Phys.\ Rev.}\ D {\bf40} 3982; Kucha\v{r} K V and Ryan M P 1986
``Can minisuperspace quantization be justified?, in {\it
Gravitational Collapse and Relativity}, edited by H. Sato and
Nakamura (World Scientific, Singapore)

\bibitem{time} Butterfield J and Isham C J 1999
``On the emergence of time in quantum gravity," {\tt
arXiv:gr-qc/9901024}; Kucha\v{r} K V 1992 ``Time and interpretations
of quantum gravity," in G Kunstatter et al {\sl General relativity
and Relativistic Astrophysics\/} (World Scientific 1992) 211--314

\bibitem{EQG} Hawking S W 1979
``The path integral approach to quantum gravity,''
in S W Hawking and W Israel {\sl General Relativity: An Einstein Centenary
Survey\/} (Cambridge University Press) 746--789;
Hartle J B and Hawking S W 1983
``Wave function of the universe,'' {\sl Phys.\ Rev.}\ D {\bf28} 2960;
Marolf D 1996 ``Path integrals and instantons in quantum gravity,''
{\sl Phys.\ Rev.}\ D {\bf53} 6979, {\tt arXiv:gr-qc/9602019}

\bibitem{CH} For a recent account, see Halliwell J J 2002
``The interpretation of quantum cosmology and the problem of time"
{\tt arXiv:gr-qc/0208018}

\bibitem{RAQ} Marolf D 1995
``Quantum observables and recollapsing dynamics,''
{\sl Class.\ Quantum Grav.}\ {\bf12} 1199,
{\tt arXiv:gr-qc/9404053};
``Almost ideal clocks in quantum cosmology: A brief derivation of
time,'' {\sl Class.\ Quantum Grav.}\ {\bf12} 2469,
{\tt arXiv:gr-qc/9412016}

\bibitem{abc} Ashtekar A, Bombelli L and Corichi A 2004 ``Coherent
and semiclassical states for constrained systems", preprint

\bibitem{kiefer1} Kiefer C 1994
``The Semiclassical approximation to quantum gravity''
{\sl Canonical Gravity: From Classical to Quantum\/} J Ehlers and H
Friedrich eds (Springer, Berlin), {\tt arXiv:gr-qc/9312015}

\bibitem{misner} Misner C W 1972 ``Minisuperspace''
in J R Klauder, ed {\sl Magic Without Magic: J A Wheeler\/} (Freeman, San
Francisco) 441--473

\bibitem{ATU} Ashtekar A, Tate R and Uggla C 1993 ``Minisuperspaces:
Observables and quantization" {\sl Int.~J.~Mod.~Phys.}\ {\bf D2} 15--50,
{\tt arXiv:gr-qc/9302017}

\bibitem{Bianchi} Ryan M P 1972 {\sl Hamiltonian Cosmology}
(Springer-Verlag)

\bibitem{ryan} Ryan M P and Waller S M 1997 ``On the Hamiltonian
formulation of class B Bianchi cosmological models''
{\tt arXiv:gr-qc/9709012}

\bibitem{Ashworth} Ashworth M C 1998 ``Coherent state approach to
time-reparametrization invariant systems" {\sl Phys. Rev.\/} A {\bf57}
2357, {\tt arXiv:quant-ph/9611026}

\bibitem{DateSingh} Date G and Singh P 2001 ``Semi-classical states
in the context of constrained systems", {\tt arXiv:quant-ph/0109127}

\bibitem{Schiff} Schiff L I 1968 {\sl Quantum Mechanics\/} 3rd edition
(McGraw-Hill)

\bibitem{Marolf}
Marolf D 1995
``Refined algebraic quantization: Systems with a single constraint"
{\tt arXiv:gr-qc/9508015};
``Group averaging and refined algebraic quantization: Where are we
now?'' {\tt arXiv:gr-qc/0011112};
Giulini D and Marolf D 1999
``On the generality of refined algebraic quantization"
{\sl Class.\ Quantum Grav.}\ {\bf16} 2479, {\tt arXiv:gr-qc/9812024}

\bibitem{rovelli} Rovelli C 1991 ``What is observable in classical
and quantum gravity?" {\sl Class. Quantum Grav.} {\bf 8} 297;
Rovelli C 1991 ``Time in quantum gravity: a hypothesis'' {\sl
Phys.\ Rev.}\ D {\bf43} 442

\bibitem{klauder1} Klauder J 2003 ``Affine quantum gravity"
{\sl Int. J. Mod. Phys.} {\bf D12} 1769--1774, {\tt arXiv:gr-qc/0305067}

\bibitem{loll} Loll R 1997 ``Imposing det $E > 0$ in discrete quantum
gravity,'' {\sl Phys.\ Lett.}\ {\bf B399} 227,
{\tt arXiv:gr-qc/9703033}

\bibitem{klauder2} Klauder K 1986 ``Global, uniform semiclassical
approximations for quantum systems on the  half-line" {\sl Phys.
Rev.} A {\bf34} 4486--4489; Watson G and Klauder J R 2000 ``Generalized
affine coherent states: A natural framework for the quantization
of metric-like variables'' {\sl J. Math. Phys.} {\bf 41} 8072--8082,
{\tt arXiv:quant-ph/0001026}

\bibitem{kiefer2} Kiefer C 1988 ``Wave packets in minisuperspace"
{\sl Phys. Rev.} D {\bf38}  1761--1772

\bibitem{kim} Kim S P, Ji J Y, Shin H S and Soh K S 1997 ``Coherence
and emergence of classical spacetime" {\sl Phys. Rev.} D {\bf56}
3756--3758, {\tt arXiv:gr-qc/9703064}

\end{thebibliography}
\end{document}